
\magnification=\magstep1
\parskip 4pt plus2pt minus2pt
\hoffset=0.1 truecm
\voffset=-0.25 truecm
\hsize=15.5 truecm
\vsize=24.5 truecm

\def\bbbq{{\mathchoice
{\setbox0=\hbox {$\displaystyle\rm Q$}\hbox
{\raise0.15\ht0\hbox to0pt{\kern0.4\wd0\vrule height0.8\ht0\hss}\box0}}
{\setbox0=\hbox {$\textstyle\rm Q$}\hbox
{\raise0.15\ht0\hbox to0pt{\kern0.4\wd0\vrule height0.8\ht0\hss}\box0}}
{\setbox0=\hbox {$\scriptstyle\rm Q$}\hbox
{\raise0.15\ht0\hbox to0pt{\kern0.4\wd0\vrule height0.7\ht0\hss}\box0}}
{\setbox0=\hbox {$\scriptscriptstyle\rm Q$}\hbox
{\raise0.15\ht0\hbox to0pt{\kern0.4\wd0\vrule height0.7\ht0\hss}\box0}}
}}

\def\bbbr{{\rm I \! R}}

\def\bbbc{{\mathchoice
{\setbox0=\hbox {$\displaystyle\rm C$}\hbox
{\hbox to0pt{\kern0.4\wd0\vrule height0.9\ht0\hss}\box0}}
{\setbox0=\hbox {$\textstyle\rm C$}\hbox
{\hbox to0pt{\kern0.4\wd0\vrule height0.9\ht0\hss}\box0}}
{\setbox0=\hbox {$\scriptstyle\rm C$}\hbox
{\hbox to0pt{\kern0.4\wd0\vrule height0.9\ht0\hss}\box0}}
{\setbox0=\hbox {$\scriptscriptstyle\rm C$}\hbox
{\hbox to0pt{\kern0.4\wd0\vrule height0.9\ht0\hss}\box0}}
}}

\font\fivesans=cmss10 at 5pt
\font\sevensans=cmss10 at 7pt
\font\tensans=cmss10
\newfam\sansfam
\textfont\sansfam=\tensans\scriptfont\sansfam=\sevensans
\scriptscriptfont\sansfam=\fivesans
\def\sans{\fam\sansfam\tensans}
\def\bbbz{{\mathchoice {\hbox{$\sans\textstyle Z\kern-0.4em Z$}}
{\hbox{$\sans\textstyle Z\kern-0.4em Z$}}
{\hbox{$\sans\scriptstyle Z\kern-0.3em Z$}}
{\hbox{$\sans\scriptscriptstyle Z\kern-0.2em Z$}}}}


\font \bigbf=cmbx10 scaled \magstep2

\def\slash#1{#1\kern-0.65em /}
\def\dirac{{\raise0.09em\hbox{/}}\kern-0.58em\partial}
\def\Dirac{{\raise0.09em\hbox{/}}\kern-0.69em D}

\def\kbar{{\mathchar'26\mkern-9muk}}




\vglue 1.5cm
\centerline {\bigbf Linear Connections on Fuzzy Manifolds}
\vskip 1.5cm

\centerline {\bf J. Madore}
\medskip
\centerline {\it Laboratoire de Physique Th\'eorique et Hautes
Energies\footnote{*}{\it Laboratoire associ\'e au CNRS {\rm URA D0063}}}
\centerline {\it Universit\'e de Paris-Sud, B\^at. 211,  \ F-91405 ORSAY}

\vskip 2cm
\noindent
{\bf Abstract:} \ Linear connections are introduced on a series of
noncommutative geometries which have commutative limits.
Quasicommutative corrections are calculated.

\vfill
\noindent
LPTHE Orsay 95/42; ESI Vienna 235
\medskip
\noindent
\bigskip
\eject

\beginsection 1 Introduction and Motivation

It is possible that the representation of space-time by a differential
manifold is only valid at length scales larger than some fundamental
length and that on smaller scales the manifold must be replaced by
something more fundamental. One alternative is a noncommutative
geometry.  If a coherent description could be found for the structure of
space-time which was pointless on small length scales, then the
ultraviolet divergences of quantum field theory could be eliminated. In
fact the elimination of these divergencies is equivalent to
course-graining the structure of space-time over small length scales; if
an ultraviolet cut-off $\Lambda$ is used then the theory does not see
length scales smaller than $\Lambda^{-1}$.  It is also believed that the
gravitational field could serve as a universal regulator, a point of
view which can be made compatible with noncommutative geometry by
supposing that there is an intimate connection between (classical and/or
quantum) gravity and the noncommutative structure of space-time. To
compare the two it is necessary to have a valid definition of a linear
connection in noncommutative geometry. There have been several examples
given of differential calculi on noncommutative geometries (Connes 1986,
Dubois-Violette 1988, Wess \& Zumino 1990).  Recently a general
definition of the noncommutative equivalent of a linear connection has
been proposed in noncommutative geometry which makes full use of the
bimodule structure of the space of 1-forms (Dubois-Violette \& Michor
1995, Mourad 1995). It has been applied to the quantum plane
(Dubois-Violette {\it et al.} 1995) and to matrix geometries (Madore
{\it et al.} 1995).

A differential manifold can always be imbedded in a flat euclidean space
of sufficiently high dimension and a linear (metric) connection on the
manifold can be considered as defined by the imbedding in terms of the
standard flat connection in the enveloping space. We shall show that
noncommutative approximations to a large class of differential manifolds
can be obtained by a similar procedure and corresponding linear
connections can be constructed as a restriction of the unique metric
connection on the enveloping matrix geometry. In the limit, when the
length parameter which determines the noncommutativity tends to zero,
first-order corrections to the commutative linear connection can
be calculated. It is these terms which must eventually be compared with
the quasiclassical corrections to the connection in quantum gravity.

Some basic formulae from previous articles are given in this Section and
in Section~2 a basic universal linear connection is introduced from
which linear connections can be constructed in a way similar to that in
which connections can be induced on an ordinary manifold when it is
imbedded in a flat space of higher dimension. The quasicommutative limit
is considered in Section~3.

Let $V$ be a differential manifold and ${\cal C}(V)$ the algebra of
smooth functions on $V$. For simplicity we suppose $V$ to be
parallelizable and we choose $\theta^\alpha$ to be a globally defined
moving frame on $V$. Let $(\Omega^*(V), d)$ be the ordinary differential
calculus on $V$. A linear connection on $V$ can be defined as a
connection on the cotangent bundle to $V$. It can be characterized as a
linear map
$$
\Omega^1(V) \buildrel D \over \rightarrow
\Omega^1(V) \otimes_{{\cal C}(V)} \Omega^1(V)                     \eqno(1.1)
$$
which satisfies the condition
$$
D (f \xi) =  df \otimes \xi + f D\xi                              \eqno(1.2)
$$
for arbitrary $f \in {\cal C}(V)$ and $\xi \in \Omega^1(V)$.

The connection form $\omega^\alpha{}_\beta$ is defined in terms of the
covariant derivative of the moving frame:
$$
D\theta^\alpha = -\omega^\alpha{}_\beta \otimes \theta^\beta.     \eqno(1.3)
$$
Let $\pi$ be the projection of
$\Omega^1(V) \otimes_{{\cal C}(V)} \Omega^1(V)$ onto $\Omega^2(V)$.
The torsion form $\Theta^\alpha$ can be defined as
$$
\Theta^\alpha = (d - \pi \circ D)\theta^\alpha.                   \eqno(1.4)
$$
The module $\Omega^1(V)$ has a natural structure as a right
${\cal C}(V)$-module and the corresponding condition equivalent to (1.2)
is determined using the fact that ${\cal C}(V)$ is a commutative algebra:
$$
D (\xi f) =  D (f \xi).                                            \eqno(1.5)
$$

By extension, a linear connection over a general noncommutative algebra
${\cal A}$ with a  differential calculus $(\Omega^*({\cal A}),d)$ can be
defined as a linear map
$$
\Omega^1({\cal A}) \buildrel D \over \rightarrow
\Omega^1({\cal A}) \otimes_{{\cal A}} \Omega^1({\cal A})           \eqno(1.6)
$$
which satisfies the condition (1.2) for arbitrary $f \in {\cal A}$ and
$\xi \in \Omega^1({\cal A})$.  The module $\Omega^1({\cal A})$ has again
a natural structure as a right ${\cal A}$-module but in the
noncommutative case it is impossible in general to consistently impose
the condition (1.5) and a substitute must be found. We must decide how
it is appropriate to define $D(\xi f)$ in terms of $D (\xi)$ and $df$.
It has been proposed (Mourad 1995, Dubois-Violette \& Michor 1995) to
introduce as part of the definition of a linear connection a map
$\sigma$ of $\Omega^1({\cal A})\otimes_{\cal A} \Omega^1({\cal A})$ into
itself and to define $D(\xi f)$ by the equation
$$
D(\xi f) = \sigma (\xi \otimes df) + (D\xi) f.                     \eqno(1.7)
$$
If the algebra is commutative this is equivalent to (1.5).  The
curvature $R$ can be defined as the map
$$
\Omega^1({\cal A}) \buildrel R \over \rightarrow
\Omega^2({\cal A}) \otimes_{{\cal A}} \Omega^1({\cal A})           \eqno(1.8)
$$
given, in the case that the torsion vanishes, by
$R = (\pi \otimes 1) \circ D^2$.

A metric $g$ on $V$ can be defined as a ${\cal C}(V)$-linear, symmetric
map of $\Omega^1(V) \otimes_{\cal C} \Omega^1(V)$ into ${\cal C}(V)$.
This definition makes sense if one replaces ${\cal C}(V)$ by an algebra
${\cal A}$ and $\Omega^1(V)$ by any differential calculus
$\Omega^1({\cal A})$ over ${\cal A}$. By analogy with the commutative
case we shall say that the covariant derivative (1.6) is metric if
$(1 \otimes g) \circ D = d \circ g$.

We shall use the conventions that lower-case Greek indices take the
values from 1 to $d$, lower-case Latin indices at the beginning of the
alphabet take the values from 1 to $m^2-1$ and the lower-case Latin
indices from $p$ to the end of the alphabet take the values from $1$ to
$n^2-1$. The integers $d,m,n$ satisfy the inequalities
$$
d < m^2-1, \qquad m < n.
$$

\beginsection 3 Induced Linear Connections

Noncommutative geometry is based on the fact that one can formulate
(Koszul 1960) much of the ordinary differential geometry of a manifold
in terms of the algebra of smooth functions defined on it. It is
possible to define a finite noncommutative geometry based on derivations
by replacing this algebra by the algebra $M_n$ of $n\times n$ complex
matrices (Dubois-Violette {\it et al.} 1989, 1990). Since $M_n$ is of
finite dimension as a vector space, all calculations reduce to pure
algebra.  Matrix geometry is interesting in being similar is certain
aspects to the ordinary geometry of compact Lie groups; it constitutes a
transition to the more abstract formalism of general noncommutative
geometry (Connes 1986, 1994). Our notation is that of Dubois-Violette
{\it et al.} (1989). We first recall some important formulae.

Let $\lambda_r$, for $1 \leq r\leq n^2-1$, be an anti-hermitian basis of
the Lie algebra of the special unitary group $SU_n$ in $n$ dimensions.
The $\lambda_r$ generate $M_n$ as an algebra and the derivations
$e_r = {\rm ad}\,\lambda_r$ form a basis for the Lie algebra of
derivations ${\rm Der}(M_n)$ of $M_n$. In order for the derivations to
have the correct dimensions we must introduce a mass parameter $\mu$
and replace $\lambda_r$ by $\mu \lambda_r$. We shall set $\mu = 1$.
We define $df$ for $f \in M_n$ by
$$
df(e_r) = e_r(f).                                              \eqno(2.1)
$$
In particular
$$
d\lambda^r(e_s) = - C^r{}_{st} \lambda^t.                      \eqno(2.2)
$$
We raise and lower indices with the Killing metric $g_{rs}$ of $SU_n$
and we use the Einstein summation convention.

We define the set of 1-forms $\Omega^1(M_n)$ to be the set of all
elements of the form $fdg$ with $f$ and $g$ in $M_n$.
The set of all differential forms is a differential algebra
$\Omega^*(M_n)$. The couple $(\Omega^*(M_n), d)$ is a differential
calculus over $M_n$.  There is a convenient system of generators of
$\Omega^1(M_n)$ as a left- or right-module completely characterized by
the equations
$$
\theta^r(e_s) = \delta^r_s.                                     \eqno(2.3)
$$
The $\theta^r$ are related to the $d\lambda^r$ by the equations
$$
d\lambda^r = C^r{}_{st}\, \lambda^s \theta^t, \qquad
\theta^r = \lambda_s \lambda^r d\lambda^s.                      \eqno(2.4)
$$
The $\theta^r$ satisfy the same structure equations as the components of
the Maurer-Cartan form on the special unitary group $SU_n$:
$$
d\theta^r =  -{1 \over 2} C^r{}_{st} \, \theta^s \theta^t.      \eqno(2.5)
$$
The product on the right-hand side of this formula is the product in
$\Omega^*(M_n)$. We shall refer to the $\theta^r$ as a frame or
Stehbein. If we define $\theta = - \lambda_r \theta^r$ we can write the
differential $df$ of an element $f \in \Omega^0(M_n)$ as a commutator:
$$
df = - [\theta, f].                                             \eqno(2.6)
$$

{}From (2.5) we see that the linear connection defined by
$$
D\theta^r = - \omega^r{}_s \otimes \theta^s, \qquad
\omega^r{}_s = -{1\over 2}  C^r{}_{st} \,\theta^t                 \eqno(2.7)
$$
has vanishing torsion. With this connection the geometry of $M_n$ looks
like the invariant geometry of the group $SU_n$.  Since the elements of
the algebra commute with the frame $\theta^r$, we can define $D$ on all
of $\Omega^*(M_n)$ using (1.2) or (1.7). The map $\sigma$ is given by
$$
\sigma (\theta^r \otimes \theta^s) = \theta^s \otimes \theta^r.   \eqno(2.8)
$$
{}From the formula (1.8) we see that
$R(\theta^r) = - \Omega^r{}_s \otimes \theta^s$ where  the curvature
2-form $\Omega^r{}_s$ is given by
$$
\Omega^r{}_s = {1\over 8} C^r{}_{st} C^t{}_{pq} \theta^p \theta^q.
                                                                  \eqno(2.9)
$$

{}From Equation (2.4) we find that $D(d\lambda^r)$ is given by
$$
D(d\lambda^r) = C^r{}_{st} \big(d\lambda^s \otimes  \theta^t
 - {1\over 2} \lambda^s C^t{}_{pq} \theta^p \otimes \theta^q\big).
$$
A short calculation yields
$$
D(d\lambda^r) =
- {1\over 2} C^r{}_{s(p} C^s{}_{q)t} \lambda^t \theta^p \otimes \theta^q.
                                                                 \eqno(2.10)
$$
{}From this formula it is obvious also that the torsion vanishes.

The connection (2.7) is the unique torsion-free metric connection on
$\Omega^1(M_n)$ (Madore {\it et al.} 1995). It has been used
(Dubois-Violette {\it et al.} 1989, Madore 1990, Madore \& Mourad 1993,
1994, Madore 1995) in the construction of noncommutative generalizations
of Kaluza-Klein theories.

Let $\{\lambda^\alpha\}$ be a set of $d$ matrices which
generate $M_n$ as an algebra and which are algebraically independent. By
this we mean that the $\lambda^\alpha$ do not satisfy any polynomial
relation of order $p$ with $p << n$. Since each $\lambda^r$ can be
written as a polynomial $\lambda^r = \lambda^r(\lambda^\alpha)$
in the $\lambda^\alpha$ we have
$$
d \lambda^r = A^r_\alpha (d \lambda^\alpha),                     \eqno(2.11)
$$
where $A^r_\alpha (d \lambda^\alpha)$ is a polynomial in
$\lambda^\alpha$ and $d\lambda^\alpha$ which is linear in the latter.
Since the $\lambda^\alpha$ generate $M_n$ it follows that the equations
$e_\alpha f = 0$ can have only $f \propto 1$ as solutions.  The algebraic
independence implies that there is no relation of the form
$$
A^\alpha_\beta (d \lambda^\beta) = 0,                            \eqno(2.12)
$$
with $A^\alpha_\beta (d \lambda^\beta)$ a polynomial of order $p-1$
in the $\lambda^\beta$.

Each choice of $\{\lambda^\alpha\}$ defines $M_n$ as a $n^2$-dimensional
approximation to the algebra of functions on a $d$-dimensional
submanifold $V$ of $\bbbr^{n^2-1}$. Let $\Omega^*_{\cal C}$ be the
associated differential calculus.  We shall argue in the next section
that a differential subalgebra of $\Omega^*_{\cal C}$ has a limit as
$n \rightarrow \infty$ which can be considered as the de~Rham differential
calculus over $V$.

{}From (2.10) we have
$$
D(d\lambda^\alpha) = - {1\over 2} C^\alpha{}_{s(p} C^s{}_{q)t}
\lambda^t \theta^p \otimes \theta^q.                               \eqno(2.13)
$$
{}From (2.4) each $\theta^r$ on the right-hand side of this equation can
in turn be expressed in terms of the $d\lambda^\alpha$:
$$
\theta^r = \lambda_s(\lambda^\alpha) \lambda^r(\lambda^\alpha)
A^s_\beta (d \lambda^\beta).                                      \eqno(2.14)
$$
Equations (2.13) and (2.14) define a covariant derivative on the
differential calculus $\Omega^*_{\cal C}$. For finite $n$ it is a
restriction of (2.7). By construction it satisfies the Leibniz rules
(1.2) and (1.7). The right-hand side however cannot be written in the
form (1.3); there is no corresponding connection form in general. The
map $\sigma$, which is given on $\theta^r \otimes \theta^s$ by the
simple expression (2.8), becomes very complicated when defined on
$d\lambda^\alpha \otimes d\lambda^\beta$.

\beginsection 3 Fuzzy manifolds

To discuss the commutative limit it is convenient to change the
normalization of the generators $\lambda^\alpha$. Recall that the
$\lambda^\alpha$ have the dimensions of mass.  We introduce the
parameter $\kbar$ with the dimensions of (length)$^2$ and define
`coordinates' $x^\alpha$ by
$$
x^\alpha = i \kbar \lambda^\alpha.                               \eqno(3.1)
$$
We define matrices $L^{\alpha\beta}$ by the equations
$$
[x^\alpha , x^\beta ] = i \kbar L^{\alpha\beta}.                 \eqno(3.2)
$$

By our assumption the $L^{\alpha\beta}$ can be expressed as polynomials
in the $x^\alpha$, normally of order $n$. By taking higher-order
commutators of the $x^\alpha$ the algebra will eventually close as a Lie
algebra to form an irreducible $n$-dimensional representation of the Lie
algebra of $SU_m$ for some $m \leq n$.  By assumption $m^2-1-d$ must be
at least as large as the number of Casimir relations of $SU_m$.  We
shall assume that $m << n$.  Let $x^a$ be the extended set of matrices:
$$
\{ x^a \} =
\{ x^\alpha, L^{\alpha\beta}, [x^\alpha, L^{\beta\gamma}],\dots \}
$$
Globally the limit manifold $V$ will be then a submanifold of the sphere
of some radius $r$ in $\bbbr^{m^2-1}$.  A metric on it would necessarily
have euclidean signature. We shall have the relation
$$
\kbar \sim {r^2 \over n},                                        \eqno(3.3)
$$
and so $\kbar \rightarrow 0$ as $n \rightarrow \infty$. This is the
commutative limit.

For each $l \geq 1$ let ${\cal C}_l$ be the vector space of
$l^{\rm th}$-order symmetric polynomials in the $x^\alpha$ and
${\cal L}_l$ the vector space of $l^{\rm th}$-order symmetric
polynomials in the $x^a$.  Then we have
$$
{\cal C}_l \subset {\cal C}_{l+1}, \qquad
{\cal L}_l \subset {\cal L}_{l+1},
$$
and the set $\{{\cal L}_l\}$ is a filtration of $M_n$:
$$
\bigcup_l {\cal C}_l \subseteq M_n, \qquad
\bigcup_l {\cal L}_l = M_n.                                      \eqno(3.4)
$$
For fixed $l$ the set ${\cal C}_l$ tends to the set of
$l^{\rm th}$-order polynomials in the $x^\alpha$ in the limit
$n \rightarrow \infty$.  We shall refer to the algebra
$M_n$ with the set of $\{{\cal C}_l\}$ as a fuzzy manifold.
The $\{{\cal C}_l\}$ do not form a graded algebra but from the
definition of the $\{{\cal L}_l\}$ we have
$$
{\cal C}_k {\cal C}_l \subset {\cal C}_{k+l} + \kbar {\cal L}_{k+l-1}.
$$

A specific example is the fuzzy 2-sphere (Madore 1992). Consider
$\bbbr^3$ with coordinates $x^a$, $1\leq a \leq 3$, and euclidean
metric $g_{ab} = \delta_{ab}$. Let $V$ be the sphere $S^2$ defined by
$$
g_{ab} x^a x^b = r^2.                                           \eqno(3.5)
$$
Consider the algebra ${\cal P}$ of polynomials in the $x^a$ and let
${\cal I}$ be the ideal generated by the relation (3.5). That is,
${\cal I}$ consists of elements of ${\cal P}$ with
$g_{ab} x^a x^b - r^2$ as factor. Then the quotient
algebra ${\cal A} = {\cal P}/{\cal I}$ is dense in the algebra
${\cal C}(S^2)$. Any element of ${\cal A}$ can be represented as a
finite multipole expansion of the form
$$
f( x^a) = f_0 + f_a x^a + {1\over 2} f_{ab} x^a x^b + \cdots,   \eqno(3.6)
$$
where the $f_{a_1 \dots a_i}$ are completely symmetric and trace-free.
We obtain a vector space of dimension $n^2$ if we consider only
polynomials of order $n-1$. We can redefine the product of the
$x^a$ to make this vector space into the algebra of $n \times n$
matrices.

Suppose that we suppress the terms $n^{\rm th}$ order in the expansion
(3.6) of every function $f$. The resulting set is a vector space
${\cal A}_n$ of dimension $n^2$. We can introduce a new product in the
$ x^a$ which will make it into the algebra $M_n$.
We make the identification
$$
x^a = \kappa J^a                                                 \eqno(3.7)
$$
where the $J^a$ generate the $n$-dimensional irreducible representation
of the Lie algebra of $SU_2$ with $[J_a,J_b] = i\epsilon_{abc}J^c$.
Since the $J^a$ satisfy the quadratic Casimir relation
$J_a J^a = (n^2-1)/4$ the parameter $\kappa$ must be related to $r$ by
the equation $4r^2 = (n^2-1)\kappa^2$. Introduce the constant
$$
\kbar = \kappa r.                                               \eqno(3.8)
$$
The $x^a$ satisfy the commutation relations
$$
[x_a , x_b ] = i \kbar C^c{}_{ab} x_c,  \qquad
C_{abc} = r^{-1} \epsilon_{abc}.                                \eqno(3.9)
$$
The two length scales $r$ and $\kbar$ are related through the integer
$n$:
$$
4r^4 = (n^2-1)\kbar^2.                                          \eqno(3.10)
$$
In particular (3.3) is satisfied. The space ${\cal L}_l$ is the space of
symmetric polynomials of order $l$ in the $x^a$. Define $x^\alpha$ as
the first two of the $x^a$. Then $L^{12} = r^{-1} x^3$. Because of
the Casimir relation we have
$$
\bigcup_l {\cal C}_l = \bigcup_l {\cal L}_l = M_n.
$$
For $n >> l$ ${\cal L}_l$ can be identified as the space of polynomials
of order $l$ on $S^2$ and ${\cal C}_l$ as the space of polynomials of
order $l$ on the coordinate patch.

The fuzzy sphere with three generators is not a good example for the
construction of linear connections since the limit manifold is not
parallelizable. Global frames must be constructed on the $U_1$
bundle $S^3$ over $S^2$. From them connections can be constructed on
$S^2$ using a Kaluza-Klein-type decomposition. (Grosse \& Madore 1991).
A more convenient example is obtained by taking only two generators. It
is known (Weyl 1931) that the algebra $M_n$ can be generated by two
matrices $u$ and $v$ which satisfy the relations
$$
u^n = 1, \qquad v^n =1, \qquad uv = qvu, \qquad q= e^{2\pi i/n}.
$$
The space ${\cal C}_l$ becomes then the space of symmetric polynomials
of order $l$ in $u$ and $v$. For $n >> l$ it can be identified as the
space of polynomials of order $l$ on the torus.

One sees from these two examples that the structure of the limit
manifold is determined by the filtration. The dimension of the manifold
is encoded in the dimension of ${\cal C}_1$.  The manifolds differ in
global topology because the vector spaces ${\cal C}_l$ differ. A
polynomial in the $x^\alpha$ of order $l$, with $n >> l$, can of course be
always written as a polynomial in $u$ and $v$ but will then in general
be of order $n$.  The transformation in no way respects the filtration.
This corresponds to the fact that a map from the torus onto the sphere
is necessarily singular. A physical theory expressed in terms of the
matrix approximation would detect the difference between the topologies
through the dependence of the action on the derivations
$e^\alpha = {\rm ad}\, x^\alpha$.

Let $\{x^\alpha\}$ be an arbitrary subset of generators of $M_n$. If we
rewrite (2.11) in terms of $x^\alpha$ we see that in the commutative
limit
$$
A^r_\alpha(dx^\alpha) =
{\partial x^r \over \partial x^\alpha} dx^\alpha + o(\kbar).
$$
This gives the differential of an arbitrary function in terms of the
differential of the coordinates.  The forms $\theta^r$ are singular in
the limit $\kbar \rightarrow 0$ (Madore 1992). No conclusions can be
drawn directly from Equation~(2.13) concerning this limit unless (2.14)
is used first to eliminate the $\theta^r$.

Consider the 1-form $[f, dg]$. It satisfies
$$
[f, dg] (X) = [f, Xg].                                            \eqno(3.11)
$$
In the limit $\kbar \rightarrow 0$ define a Poisson bracket $\{f, g\}$
on $V$ by
$$
i\kbar \{f, g\} = [f, g].                                         \eqno(3.12)
$$
By taking the limit of (3.11) we can define the extension $\{f, dg\}$ by
$$
\{f, dg\} (X) = \{f, Xg\}.                                        \eqno(3.13)
$$
It is obvious that $\{f, dg\}$ is not an element of $\Omega^1(V)$.  It
is a $\bbbc$-linear map of the derivations into the functions but it
cannot be ${\cal C}(V)$-linear, because Poisson vector fields do not
form a ${\cal C}$-module.  The 1-form defined by (3.11) contains a term
of order $\kbar$ which cannot be approximated by an element of
$\Omega^1(V)$. Define $\Omega^1_{\cal C}(V)$ to be the 1-forms of a new
differential calculus on $V$ defined by (3.13). We have seen then that
$$
\Omega^1_{\cal C}(V) \neq \Omega^1(V).                           \eqno(3.14)
$$
In a sense the left-hand side is smaller since it is only defined on
Poisson vector fields. However since
$$
d\{f, g\} = \{df, g\} + \{f, dg\}                                \eqno(3.15)
$$
every element of $\Omega^1(V)$ defines by restriction an element of
$\Omega^1_{\cal C}(V)$. So in a sense the left-hand side is larger.
The map $d$ of $\Omega^1_{\cal C}(V)$ into $\Omega^2_{\cal C}(V)$ is
defined by $d \{f, dg\} = \{df, dg\}$ with
$$
\{df, dg\}(X,Y) = \{Xf, Yg\} - \{Yf, Xg\}.
$$
The image is also not ${\cal C}(V)$-linear and would not coincide
with the bracket of 1-forms defined, for example, by Koszul (1985).

We define the element $dx^{ab}$ of $\Omega^1_{\cal C}(V)$ as
$$
dx^{ab} = \{x^\alpha, dx^\beta\}.                                \eqno(3.16)
$$
We can write the induced connection in the quasicommutative limit in the
form
$$
\eqalign{
&D(dx^\alpha) =
-\Gamma^\alpha{}_{\beta\gamma} dx^\beta \otimes dx^\gamma
- \kbar \Gamma^\alpha_{(1)} + o(\kbar^2),                         \cr
&D(dx^{\alpha\beta}) = - \Gamma^{\alpha\beta}_{(1)} + o(\kbar),
}                                                                 \eqno(3.17)
$$
where
$$
\Gamma^\alpha_{(1)} =
\Gamma_L{}^\alpha{}_{\beta\gamma\delta}
dx^{\beta\gamma} \otimes dx^\delta +
\Gamma_R{}^\alpha{}_{\beta\gamma\delta}
dx^d \otimes dx^{\beta\gamma}.                                    \eqno(3.18)
$$
The $\Gamma_L{}^\alpha{}_{\beta\gamma\delta}$ and
$\Gamma_R{}^\alpha{}_{\beta\gamma\delta}$ can be considered as
functions on the limit manifold $V$. Although the right-hand side of
(2.13) is symmetric in $p$ and $q$, in general because of our convention
of placing all coefficients of forms to the left of the differential,
$$
\Gamma_L{}^\alpha{}_{\beta\gamma\delta} \neq
\Gamma_R{}^\alpha{}_{\beta\gamma\delta}.
$$
The right-hand side of the second equation (3.17) is an element of
$\Omega^1_{\cal C}(V) \otimes \Omega^1_{\cal C}(V)$.

We have deduced the form of the Equations~(3.17) from (2.13) and (2.14).
They depend however only on the Poisson structure, through the
differential calculus $\Omega^*_{\cal C}(V)$.  The Poisson structure is
the unique `shadow' of the original noncommutative algebra and the extra
terms on the right-hand side of (3.17) the unique `shadow' of the
noncommutative linear connection. As we have mentioned the manifolds we
can approximate in this way are compact with metrics necessarily of
euclidean signature. They are of interest in that their algebra of
functions can be approximated by algebras of finite dimension. Of more
physical relevance for relativistic physics are noncompact manifolds
which can support metrics of Minkowski signature.  The first example
along the lines indicated by the relation (3.2) was given by Snyder
(1947). See also Madore (1988, 1995). Doplicher {\it et al.} (1995) have
given an analysis of several possible noncommutative extensions of
Minkowski space within the context of relativistic quantum field theory.

\parindent=0cm
{\it Acknowledgment:}\ This research was completed while the author
was visiting at the Erwin Schr\"odinger Institut in Vienna. He would
like to thank the acting director Peter Michor for his hospitality. He
would also like to thank M. Dubois-Violette and J. Mourad for
enlightening discussions.

\beginsection References
\parskip 5pt plus 1pt

Connes A. 1986, {\it Non-Commutative Differential Geometry}, Publications
of the Inst. des Hautes Etudes Scientifique. {\bf 62} 257.

--- 1994, {\it Noncommutative Geometry}, Academic Press.

Doplicher S., Fredenhagen K., Roberts J.E. 1995 {\it Quantum Structure
of Spacetime at the Planck Scale and Quantum Fields}, Commun. Math.
Phys. (to appear).

Dubois-Violette M. 1988, {\it D\'erivations et calcul diff\'erentiel
non-commutatif}, C. R. Acad. Sci. Paris {\bf 307} S\'erie I 403.

Dubois-Violette M., Madore J. Masson T. Mourad J. 1995,
{\it Linear Connections on the Quantum Plane}, Lett. Math. Phys. (to appear).

Dubois-Violette M., Kerner R., Madore J. 1989, {\it Gauge bosons in a
noncommutative geometry}, Phys. Lett. {\bf B217} 485; {\it Classical
bosons in a noncommutative geometry}, Class. Quant. Grav. {\bf 6} 1709.

--- 1990, {\it Noncommutative differential geometry of matrix algebras},
J. Math. Phys. {\bf 31} 316.

Dubois-Violette M., Michor P. 1994, {\it D\'erivations et calcul
diff\'erentiel non-commuta\-tif II}, C. R. Acad. Sci. Paris {\bf 319}
S\'erie I 927.

Dubois-Violette M., Michor P. 1995, {\it Connections on Central
Bimodules}, Preprint LPTHE Orsay 94/100.

Grosse H., Madore J.  1991, {\it A Noncommutative Version of the
Schwinger Model}, Phys. Lett. {\bf B283} 218.

Koszul J.L. 1960, {\it Lectures on Fibre Bundles and Differential Geometry},
Tata Institute of Fundamental Research, Bombay.

Koszul J.L. 1985, {\it Crochet de Schouten-Nijenhuis et Cohomologie},
Soc. Math. de France, Ast\'erisque, p. 257.

Madore J. 1988, {\it Non-Commutative Geometry and the Spinning Particle},
Lecture given at the XI Warsaw Symposium on Elementary Particle Physics,
Kazimierz, Poland.

Madore J. 1990, {\it Modification of Kaluza-Klein Theory}, Phys. Rev.
{\bf D41} 3709.

Madore J. 1992, {\it The Fuzzy Sphere}, Class. Quant. Grav. {\bf 9} 69.

Madore J. 1995, {\it Noncommutative Differential Geometry and its
Physical Applications}, Cambridge University Press.

Madore J., Masson T., Mourad J. 1995, {\it Linear Connections on
Matrix Geometries}, Class. Quant. Grav. {\bf 12} 1429.

Madore J., Mourad J. 1993, {\it Algebraic-Kaluza-Klein Cosmology},
Class. Quant. Grav. {\bf 10} 2157.

--- 1994, {\it A Noncommutative Extension of Gravity},
Int. Jour. of Mod. Phys.~D, {\bf 3} 221.

Mourad. J. 1995, {\it Linear Connections in Non-Commutative Geometry},
Class. Quant. Grav. {\bf 12} 965.

Snyder H.S. 1947, {\it Quantized Space-Time}, Phys. Rev. {\bf 71} 38.

Wess J., Zumino B. 1990, {\it Covariant Differential Calculus on the
Quantum Hyperplane} Nucl. Phys. B (Proc. Suppl.) {\bf 18B} 302.

Weyl H. 1931, {\it The Theory of Groups and Quantum Mechanics}, Dover.

\bye